# Electrochemical Grafting of α-Naphthylmethyl Radicals to Epitaxial Graphene: A Versatile Platform to Reversibly Engineer the Band Structure and Transport Properties of Graphene**

*Santanu Sarkar, Elena Bekyarova, Robert C. Haddon**

*Centre for Nanoscale Science and Engineering, Departments of Chemistry and Chemical & Environmental Engineering, University of California, Riverside, California 92521-0403, USA. Fax: (+1) 951-827-5696. E-mail: haddon@ucr.edu*

**Abstract:** The Kolbe electrochemical oxidation strategy has been utilized to achieve an efficient quasireversible electrochemical grafting of the α-naphthylmethyl functional group to graphene. The method facilitates reversible bandgap engineering in graphene and preparation of electrochemically erasable organic dielectric films. The picture shows Raman D-band maps of both systems.

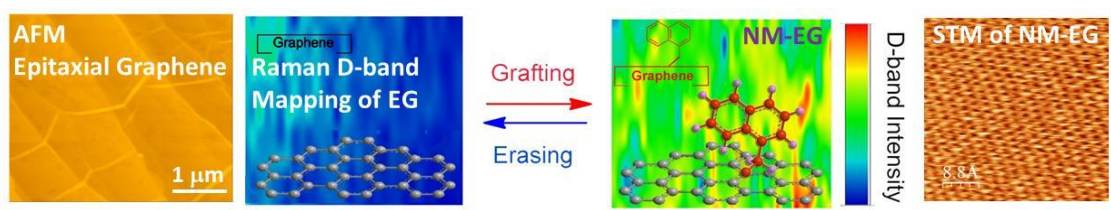

The potential of graphene as an electronic material has generated significant research activity toward the goal of engineering a band gap in this material and the application of chemistry for the modification of the electronic and magnetic structures of graphene has emerged as a promising approach.[1-8] From the chemical standpoint graphene is a particularly intriguing material;[9-12] although graphene constitutes the thermodynamic ground state of carbon and is solely comprised of sp$^2$ hybridized carbon atoms, the unique electronic structure of graphene allows it to participate in surprisingly mild reaction processes.[13, 14]

In the present communication we show that graphene readily undergoes the Kolbe reaction (eq.1),[15-17] which involves the electrochemical oxidation of carboxylates with subsequent grafting of the derived carbon radicals (Figure 1 and Scheme 1).

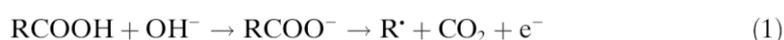

$$RCOOH + OH^- \rightarrow RCOO^- \rightarrow R^\bullet + CO_2 + e^- \qquad (1)$$

In this report we demonstrate the reversible grafting of α-naphthylmethyl groups to epitaxial graphene (EG) as a versatile approach for engineering the electronic band structure of graphene. The advantages of the Kolbe electro-oxidation in the chemical modification of graphene are: (i) reversibility of the reaction - the grafted functionality can be electrochemically erased, (ii) α-naphthylmethyl (α-NM) groups are found to offer well-ordered structural patterning on graphite surfaces,[18] and thus the resulting graphene derivative is anticipated to exhibit interesting magnetic and electronic behavior, and (iii) simplicity, versatility and efficiency of the reaction that makes possible the covalent binding of a wide variety of arylmethyl groups with appropriate substituents on the phenyl rings.[18]

The grafting of α-NM groups to the EG surface was performed by anodic oxidation of a α-naphthylacetate (Figure 1a, process **1** in Scheme 1); this process produces α-naphthylmethyl (α-NM) radicals in the vicinity of the graphene surface, which rapidly leads to the covalent attachment of the α-NM functionality to the graphene lattice via the formation of C–C bond and subsequent creation of an sp$^3$ carbon centre in the graphene lattice (processes **2** and **3** in Scheme 1).

The experiments were performed using a 5 × 5 mm$^2$ EG wafer as the working electrode immersed in a solution of α-naphthylacetic acid and n-Bu$_4$NOH in acetonitrile, to which ~0.1 M n-Bu$_4$NPF$_6$ was added as an electrolyte (see Supporting Information for details). Figures 1b, 1c, and 1d show successive cyclic voltammograms recorded at EG and HOPG electrodes; the oxidation of the α-naphthylacetate occurs at ~0.93 V vs SCE. During the derivatization of EG with α-NM groups the anodic current (oxidation peak) in the cyclic voltammetry curve vanishes almost completely after the first scan (Figure 1b, scan rate = 0.2 V.s$^{-1}$), irrespective of the concentration of the α-naphthylacetate solution. This indicates the complete passivation of the EG surface due to the attached α-NM functionality.

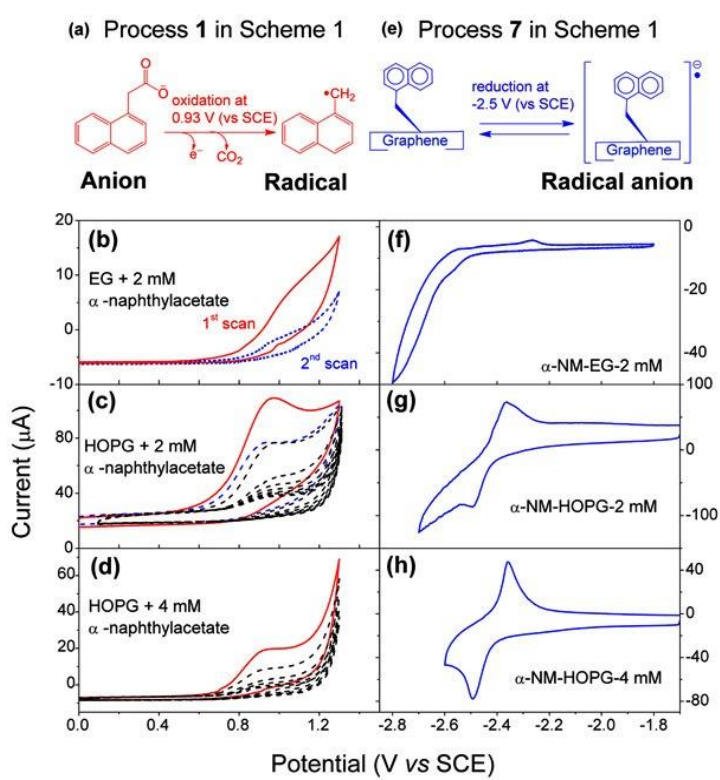

*Figure 1.* (a) Generation of an α-naphthylmethyl radical by oxidation of α-naphthylacetate. Oxidative cyclic voltammetry of (b) EG-electrode in 2 mM α-naphthylacetic acid (α-NAA), (c) HOPG in ~2 mM α-NAA, and (d) HOPG in ~4 mM α-NAA; the solutions were prepared with CH$_3$CN and contained n-Bu$_4$NOH and ~0.1 M n-Bu$_4$NPF$_6$. Solid line: first scan, dotted line: sucessive scans; scan rate = 0.2 Vs$^{-1}$. (e) Reduction of the α-naphthylmethyl group attached to graphene. Reductive cyclic voltammetry of (f) α-NM-EG and (g, h) α-NM-HOPG electrodes (derivatized using the electrochemical processes in the left frames) with ~0.1 M n-Bu$_4$NPF$_6$ in CH$_3$CN.

The efficient passivation of EG is in contrast to the passivation of HOPG, which occurs progressively and depends on the concentration of the arylacetate as illustrated in Figures 1 c and 1d. Thus, in case of a 2 mM α-naphthylacetate solution the number of cycles required for the passivation of the HOPG electrode at a scan rate of 0.2 V.s$^{-1}$ was 11 (Figure 1c), whereas 4 cycles were necessary when a 4 mM solution was used (Figure 1d). This phenomenon is attributed to the competing dimerization of the α-NM radicals, which is operative only in the presence of the less reactive graphite (HOPG) surface, and not on the graphene (EG) surface.[11, 19, 20]

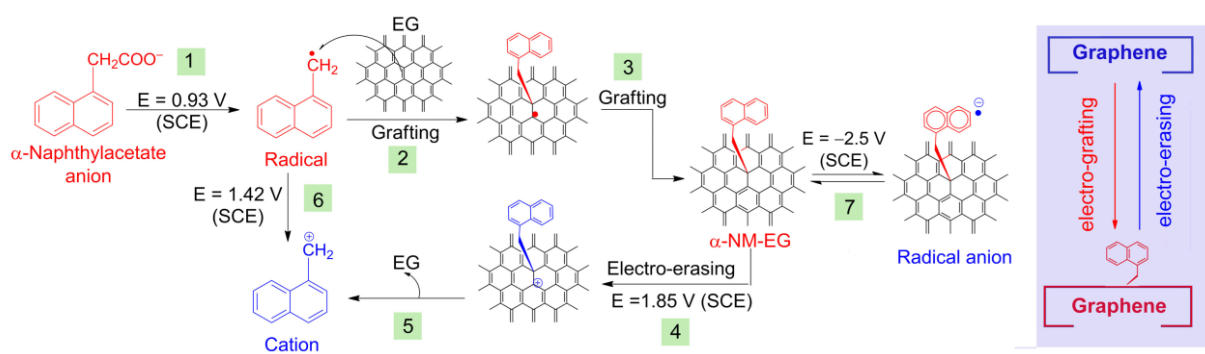

*Scheme 1.* Mechanistic pathways associated with the grafting of α-naphthylmethyl (α-NM) groups to epitaxial graphene (EG).

The covalent attachment of the α-NM radical to the epitaxial graphene (processes **2** and **3** in Scheme 1) creates a new sp$^3$ carbon center in place of an sp$^2$ carbon atom in the graphene lattice and this is readily detected by Raman spectroscopy with the development of a D-band at ~1345 cm$^{-1}$ as shown in Figure 2a. The Raman spectrum of the pristine EG sample shows the characteristic G, G', 2D and 2D' bands (Figure 2a), whereas the D, D$^*$ and D+D' bands appear in the spectra of the α-NM-EG product; the intensity of the 2D band is reduced by functionalization as observed in the addition of nitrophenyl groups to graphene.[4, 5]

The Raman intensity map of the D-band in the graphene samples is shown in Figure 2; the map of pristine EG (Figure 2b) shows that the selected area of the wafer is defect-free, whereas covalent functionalization of the same EG surface leads to the appearance of a prominent D-peak (Figure 2c).

The attachment of α-NM group to EG was further confirmed by ATR-IR spectroscopy (Figure 2d); the spectrum of α-NM-EG shows the characteristic intense band at ~792 cm$^{-1}$, which is ascribed to the in-phase C–H wagging vibrations of aryl groups and similar peaks in α-naphthylacetic acid and naphthalene appear at ~779 and 774 cm$^{-1}$ respectively (Figure S7 in Supporting Information).[21, 22]

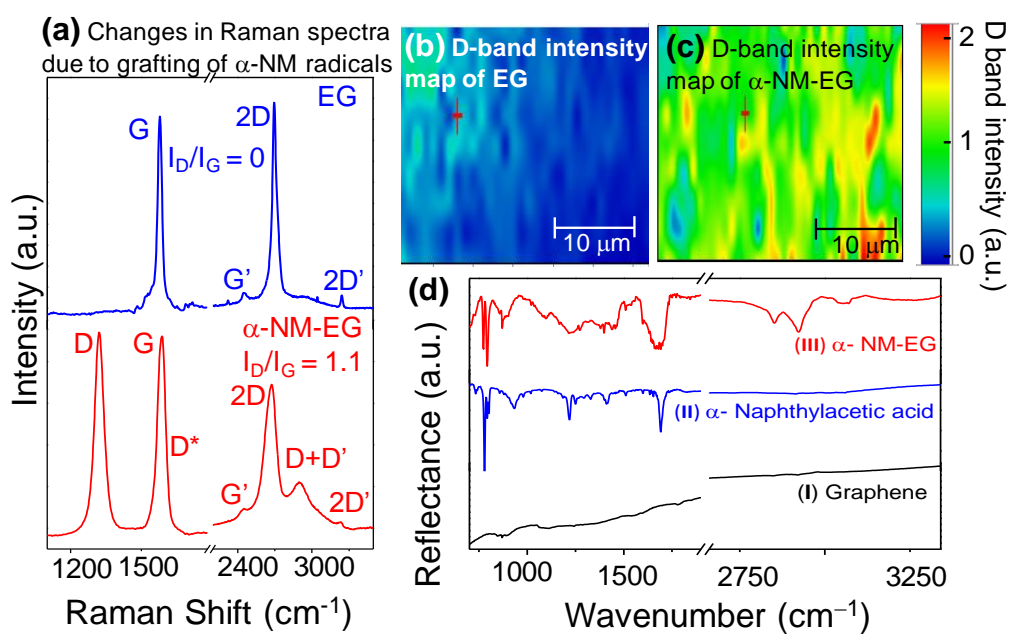

*Figure 2.* (a) Raman spectra (excitation wavelength, $\lambda_{ex}$ = 532 nm) before (EG) and after electrochemical grafting of α-naphthylmethyl group to EG (α–NM-EG). α–NM-EG was prepared by complete passivation as shown in Figure 1b. Raman intensity map of the D-band in (b) pristine EG and (c) α–NM-EG for a selected area of the wafer; (d) ATR-IR spectra of (I) pristine EG, (II) α-naphthylacetic acid, and (III) α-NM-EG.

The oxidation waves of the grafted groups (process **4** in Scheme 1) were irreversible at low scan rates in a pure electrolyte solution and these waves disappeared after the second anodic scan (Figure 3a and 3b), showing the erasure of the grafted functionalities under electro-oxidative conditions (process

**5** in Scheme 1). Thus, electro-erasing of the α-NM-EG films was achieved by running two cycles of an oxidative CV between 1 and 2.5 V vs SCE and this is illustrated in Figure 3a for α-NM-EG and Figure 3b for α-NM-HOPG.

After electrochemical erasure, the resulting EG or HOPG electrode behaved like a clean EG or HOPG electrode as may be seen by running the reductive cyclic voltammetry of the electro-erased electrodes, which are essentially featureless (Figures 3c and 3d). After electro-erasing of the α-NM-groups from the α-NM-EG electrode, the surface can be re-functionalized under the oxidative CV conditions shown in Figure 1b, and the electrode exhibited the same behaviour towards passivation by α-naphthylacetate as the pristine EG-electrode.

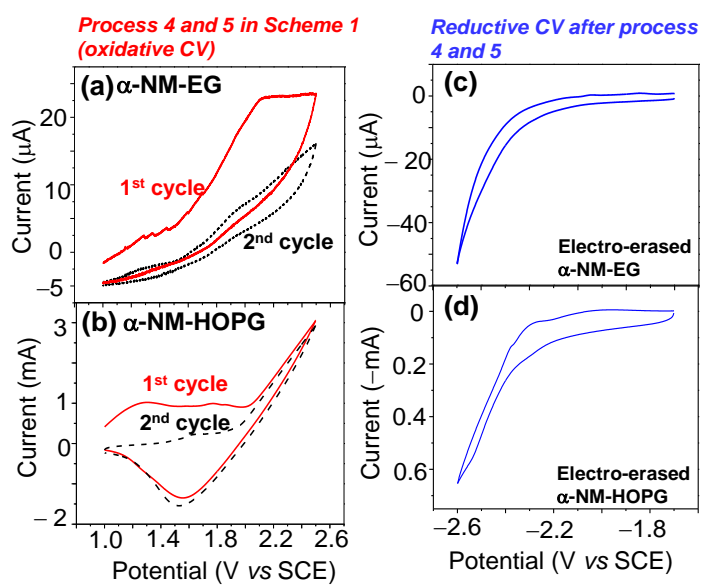

*Figure 3.* Electrochemical erasure of the α-NM-groups from (a) α-NM-EG electrode and (b) α-NM-HOPG electrode by oxidative cleavage. Reductive cyclic voltammetry of electro-erased (c) α-NM-EG and (d) α-NM-HOPG electrodes (scan rate = 0.2 Vs$^{-1}$).

The fidelity of the electro-grafting (process **2, 3**) and -erasing (process **4**, **5**) steps is apparent in the evolution of the D-band in the Raman spectrum as a function of the electrochemical treatment (Figure 4).

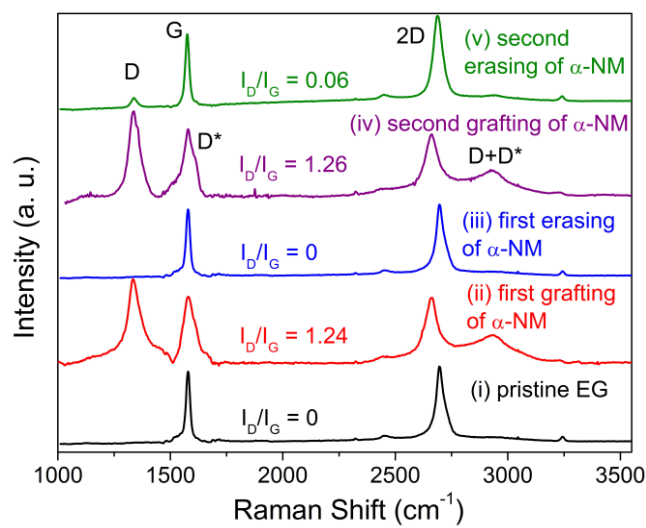

*Figure 4.* Evolution of the EG Raman spectrum (excitation wavelength, $\lambda_{ex}$ = 532 nm) following multiple electrochemical grafting and erasing steps of the α-naphthylmethyl group: (i) pristine EG, (ii) after first grafting of α-NM (α-NM-EG), (iii) after first electrochemical erasing of α-NM functional group from α-NM-EG, (iv) after second grafting of α-NM to electro-erased α-NM-EG, and (v) second electrochemical erasing.

Alternatively, the α-NM groups can be elecrochemically erased by transferring the α-NM–EG electrode to a pure electrolyte solution and setting the potential at the level of the oxidation wave. Thus potentiostatic electrolysis of the α-NM-EG electrode at 1.85 V vs SCE for 240 s in a pure electrolyte solution produces a subsequent CV which is essentially featureless, suggesting the efficient erasure of the grafted functionality and the restoration of the initial structure of the epitaxial graphene.

The derivatized surfaces were further characterized by analyzing the reductive cyclic voltammetry, which was conducted in a pure electrolyte solution. The one-electron reduction of the attached α-NM groups (Figure 1e, process **7** in Scheme 1) gives rise to a reduction wave (Figure 1f-h) and the surface coverage ($\Gamma$) of the α-NM groups can be estimated from the charge using the formula: $\Gamma = Q/nFA$, where $Q$ is the integrated area of the reduction peak (Coulombs of charge), $n$ is the number of electrons (n=1), $F$ is the Faraday constant ($9.648 \times 10^4$ C.mol$^{-1}$), and $A$ is the area of the electrode. The functionalized α-NM-EG samples, obtained by complete passivation of the EG surface in a α-naphthylacetate solution (as illustrated in Figure 1b), were found to have an approximate surface coverage of $10 \times 10^{-10}$ mol.cm$^{-2}$ (Figure 1f), which corresponds to a densely packed layer of α-NM groups. For the EG substrate the surface coverage was found to be independent of the concentration of the α-naphthylacetate, while for the HOPG functionalization the surface coverage was found to be: $4.5 \times 10^{-10}$ mol.cm$^{-2}$ (Figure 1g, 2 mM), and $9.5 \times 10^{-10}$ mol.cm$^{-2}$ (Figure 1h, 4 mM).[19, 20]

AFM images of the α-NM-EG surface, which were recorded after cleaning the substrate by ultrasonication in isopropanol, confirmed the formation of a dense layer on the EG surface (Figure 5b), which suggests that these layers may function as dielectric films. In another experiment the EG was functionalized by potentiostatic electrolysis in a 2 mM α-naphthylacetates at 0.8 V vs SCE for 2.5 seconds; a subsequent reductive CV gave a reduction peak at –2.5 V vs SCE with an area corresponding to a surface coverage of ~$0.49 \times 10^{-10}$ mol.cm$^{-2}$ (see Supporting Information for details). The AFM image confirmed a sparse coverage of the EG surface (Figure 5c). Control experiments on HOPG substrates show that the film thickness can be controlled by the applied potential and the scan duration.

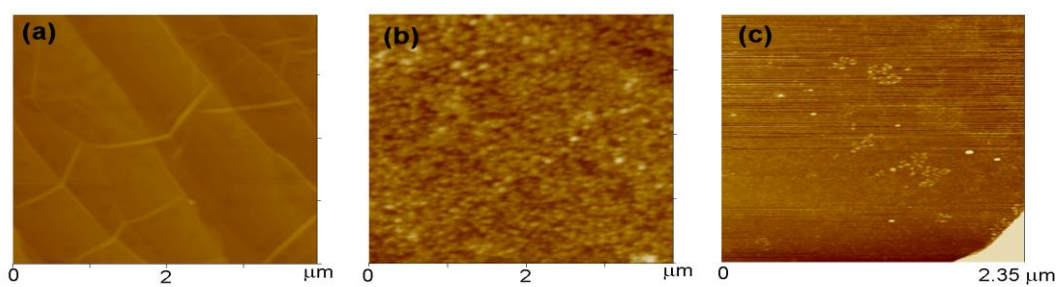

*Figure 5*. AFM images of (a) pristine EG, (b) completely passivated α-NM-EG obtained by oxidative CV runs in 2 mM α-naphthylmetyl acetate between 0 and 1.3 V

vs SCE, and (c) sparsely functionalized α-NM-EG obtained by controlled potentiostatic electrolysis of 2 mM α-naphthyl acetate at 0.8 V vs SCE for 2.5 seconds.

In conclusion, our present work demonstrates that arylmethyl groups can be grafted electrochemically to the surface of epitaxial graphene, which shows higher reactivity than HOPG, and the EG grafting was found to be independent on the concentration of the α-naphthylacetic acid used in this study (2 – 4 mM). The surface coverage of naphthylmethyl groups can be controlled by adjustment of the electrochemical conditions for functionalization of EG; the control of the layer structure and packing of the functional groups over the graphene surface is an essential issue in the development of graphene chemistry.[14, 23-25] The functionalization is readily reversed and may be repeated in a simple, efficient and reproducible manner suggesting the potential of this approach for reversible engineering of the band structure and conductivity. The versatility of this method offers new routes for the post-grafting modification and well-ordered structural patterning of graphene wafers for advanced electronic, magnetic and electro-optical applications.

**Experimental Section**

Epitaxial graphene (EG) samples, grown on single crystal SiC (0001) by vacuum graphitization, were provided by Professor Walt de Heer (Georgia Institute of Technology). All experiments were performed on the C-face of the EG. HOPG samples were obtained from Union Carbide Corporation. α-Naphthylacetic acid (FW = 186.21), tetrabutylammonium hexafluorophosphate (n-Bu$_4$NPF$_6$, FW = 387.43), tetrabutyl-ammonium hydroxide 30-hydrate (n-Bu4N$^+$OH$^-$.30H$_2$O, FW = 259.47) and acetonitrile (anhydrous, 99.9 %) were obtained from Sigma-Aldrich. Electrochemical experiments were carried out with a computer-controlled CH Instruments Electrochemical Analyzer. Raman spectra were collected in a Nicolet Almega XR Dispersive Raman microscope with a 0.7 μm spot size and 532 nm laser excitation. The ATR-IR spectra were taken using a Thermo Nicolet Nexus 670 FTIR instrument, equipped with an ATR sampling accessory.

The EG and HOPG samples for electrochemical surface functionalization reactions were fixed on a glass substrate with pre-patterned gold contacts. The graphene samples were electrically contacted with silver paint and the contacts were isolated with epoxy resin. The EG (or HOPG) substrate served as the working electrode, while the platinum (Pt) wire and saturated calomel electrode (SCE) were used as counter and reference electrodes, respectively. The solutions of α-naphthylacetate were prepared in a glove box (see SI for details). The electrochemical cell with the substrate and solution was purged with argon prior to use.

# Supporting Information for

## Electrochemical Grafting of α-Naphthylmethyl Radicals to Epitaxial Graphene: A Versatile Platform to Reversibly Engineer the Band Structure and Transport Properties of Graphene**

*Santanu Sarkar, Elena Bekyarova, Robert C. Haddon\**

*Centre for Nanoscale Science and Engineering, Departments of Chemistry and Chemical & Environmental Engineering, University of California, Riverside, California 92521-0403, USA. Fax: (+1) 951-827-5696. E-mail: haddon@ucr.edu*

**Table of Contents**

## 1. Schematics of Kolbe's electrochemical oxidation and grafting of radicals to graphene:

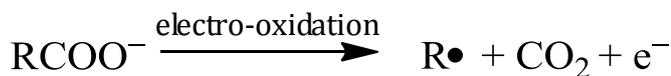

$$\text{RCOO}^- \xrightarrow{\text{electro-oxidation}} \text{R}\bullet + \text{CO}_2 + \text{e}^-$$

**Example: oxidation to α-naphthylacetate**

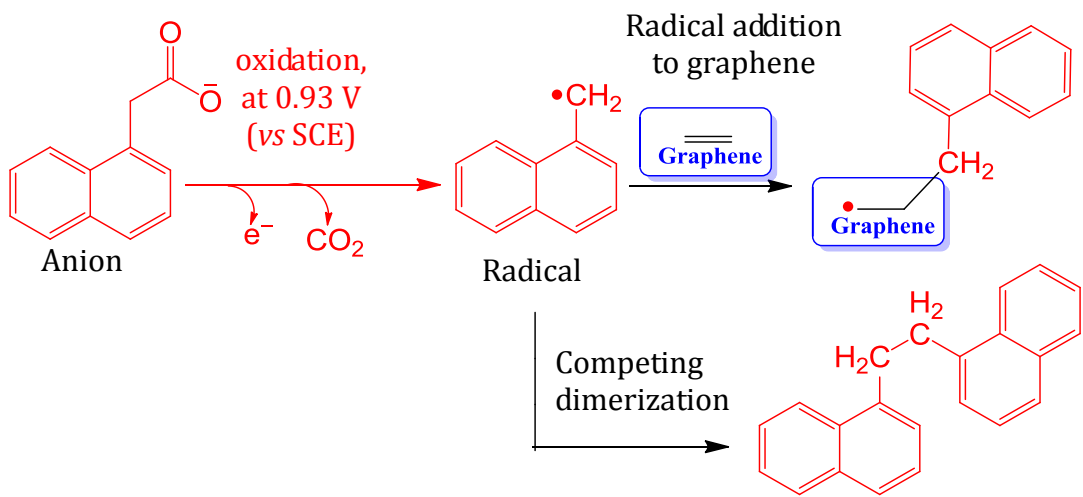

## 2. Sample preparation for electrochemistry:

Epitaxial graphene (EG) grown on the C-face of SiC substrates (Cree Inc, 4H SiC) with average dimensions of 3.5 cm × 4.5 cm were provided by Professor Walt de Heer (Georgia Tech). The EG samples were fixed on a glass slide with pre-patterned gold contacts using silver paint and all metal surfaces were sealed with epoxy (**Figure S1a**). HOPG samples for electrochemical functionalization reactions were prepared by fixing a piece of HOPG on a glass slide and contacted with silver paint; all metal surfaces were then coated with an epoxy resin to insulate them from the electrolyte solution (**Figure 1Sb**). The EG substrate (or HOPG) were connected as a working electrode and immersed in the solution for electrochemistry.

**Preparation of 4 mM and 2 mM solutions of α-naphthylacetate:** α-Naphthylacetic acid (15 mg, 0.08 mmol, F. W. = 186.21), tetrabutylammonium hydroxide 30-hydrate (21 mg, 0.08 mmol, F. W. = 259.47) and tetrabutylammonium hexafluorophosphate (775 mg, 2.06 mmol, F. W. = 387.43) were weighed in a vial inside a glove box; anhydrous acetonitrile (20 mL) was added and the resulting mixture was homogenized by shaking to obtain a ~4 mM α-naphthylacetate solution. Similarly ~2 mM solutions of α-naphthylacetate in acetonitrile was prepared using α-naphthylacetic acid (7.5 mg, 0.04 mmol), and tetrabutylammonium hydroxide 30-hydrate (11 mg, 0.04 mmol).

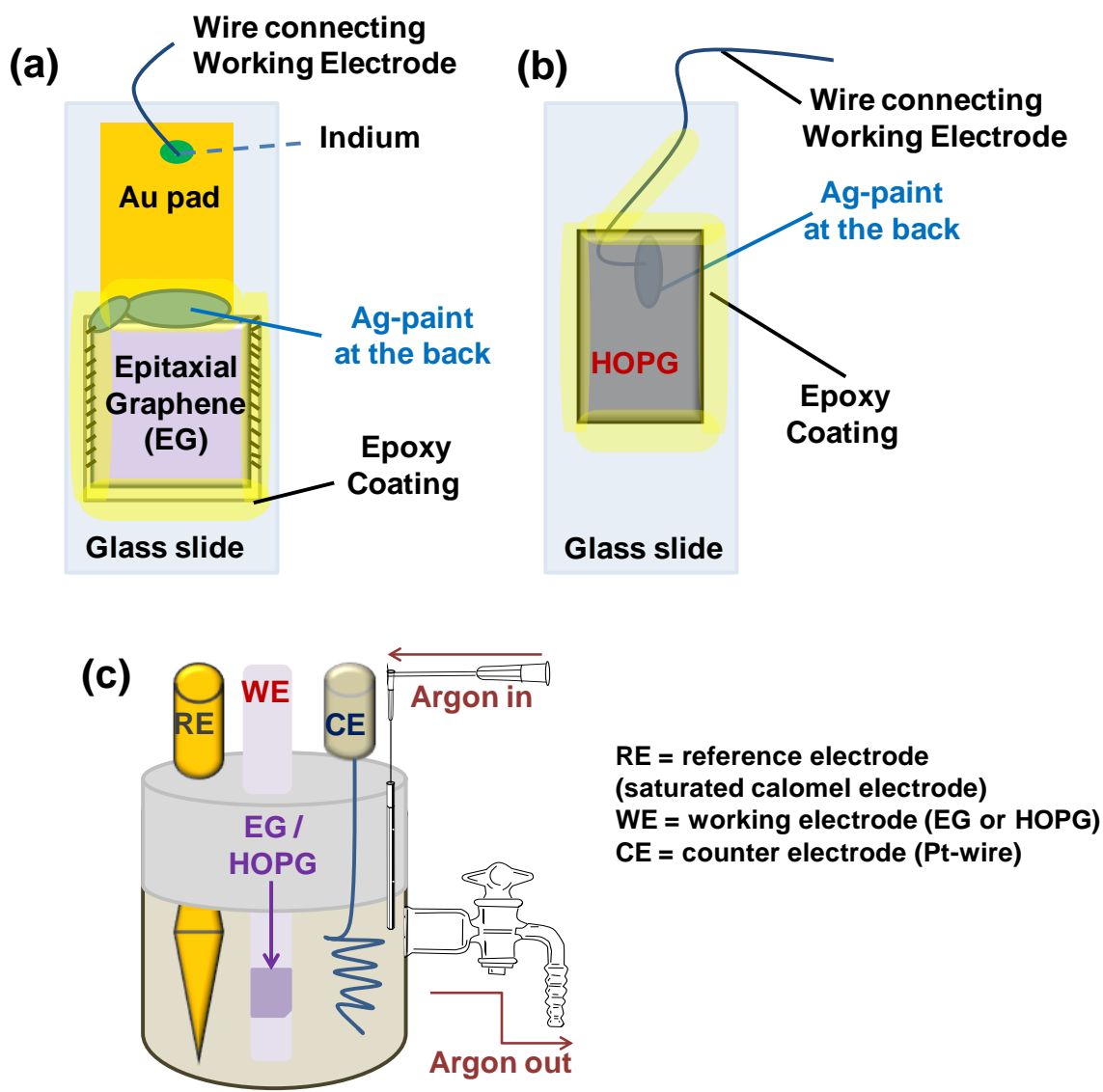

**Figure S1.** Sample preparation for electrochemical functionalization of (a) epitaxial graphene and (b) HOPG mounted on a glass substrate using Ag-paint. (c) Typical configuration of the electrochemical cell used for the generation and electro-grafting of α-naphthylmethyl radicals to epitaxial graphene and HOPG.

## 3. Electrochemical grafting of α-naphthylmethyl (α-NM) groups to epitaxial graphene (EG) and highly oriented pyrolytic graphite (HOPG):

A computer-controlled CH Instruments Electrochemical Analyzer was used to carry out the electrochemical experiments. The EG (or HOPG) substrate served as the working electrode, while the platinum (Pt) wire and saturated calomel electrode (SCE) were the counter and reference electrodes, respectively. The electrochemical cell containing the electrolyte solution with the three-electrode setup was purged with argon prior to use.

The derivatization of EG or HOPG with α-naphthylmethyl group was performed by repetitive oxidative cyclic voltammetry between 0 to 1.3 V vs SCE at a scan rate of 0.2 Vs$^{-1}$. The oxidation of α-naphthylacetate was found to occur at 0.93 V *vs* SCE (**Figure S2a**).

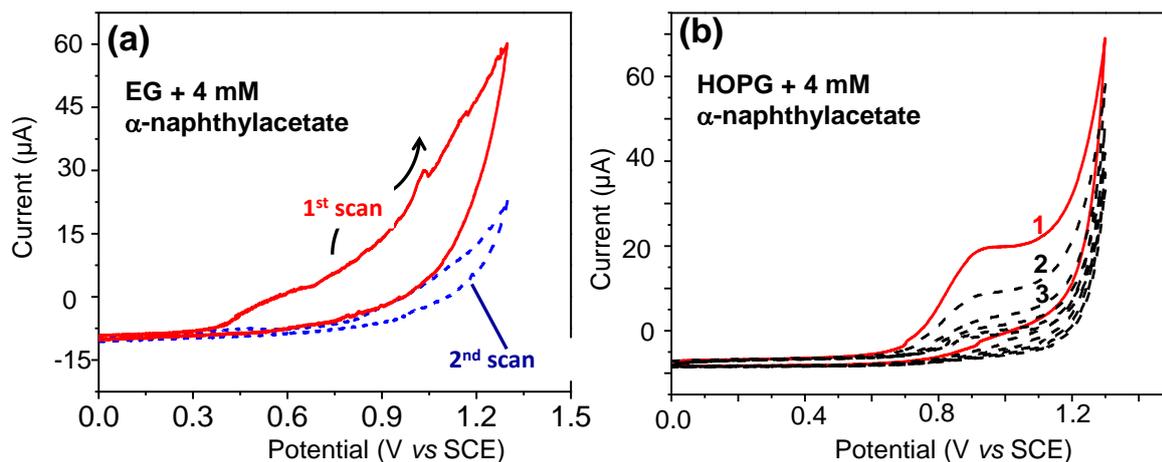

**Figure S2.** Successive cyclic voltammograms of oxidation waves recorded at (a) epitaxial graphene (EG) and (b) HOPG electrodes in presence of 4 mM α–naphthylacetate and ~0.1 M $n$-Bu$_4$NPF$_6$ in acetonitrile (CH$_3$CN) solution. The oxidation waves resulted in grafting of α-naphthylmethyl groups to the EG surface – processes 1 and 2 in Scheme 1 (see main text).

During repetitive oxidative sweeps it was observed that in case of HOPG, the number of cycles required to completely passivate the electrode (diminished anodic current) depends on the concentration of α-naphthylacetate used for the grafting process; 11 cycles (**Figure 1c**) were required for 2 mM and and 4 cycles (**Figure 1d**) for 4 mM α-naphthylacetate. This contrasts with the passivation of EG with α-NM, which has occured after the first anodic scan (**Figure 1b,** and **Figure S2a**). The EG or HOPG substrate thus derivatized (referred to as α-NM-EG or α-NM-HOPG) was taken out from the electrochemical cell and carefully rinsed with four solvents: acetonitrile, ethanol, dichloromethane and acetone and then dried under a gentle flow of argon prior to further studies.

## 4. Calculation of surface coverage of α-NM-EG and α-NM-HOPG surfaces after complete passivation with α-NM group:

The surface coverage of the grafted α-NM groups on EG (or HOPG) surface was determined by reductive cyclic voltammetry in a potential window between –1.7 and –2.7 V vs SCE at a scan rate of 0.2 Vs$^{-1}$ (**Figure S3**) in a pure electrolyte solution (~0.1 M $n$Bu$_4$NPF$_6$ in CH$_3$CN) The integrated area under the reduction peak located at about –2.5 V vs SCE gives the reduction charge for this one-electron reduction process.

<u>*Surface coverage calculation:*</u>
Surface coverage, $\Gamma = Q/(nFA)$,
where $\Theta$ = integrated area under the reduction peak of the current-time plot (in Coulomb),
n = number of electrons involved in this reduction process (n =1 in the present case),
F = $9.648 \times 10^4$ C.mol$^{-1}$, and A = surface area of the EG or HOPG surface (in cm$^2$).

(i) <u>**EG-electrode functionalized using 2 mM α-naphthylacetate solution (Figure 1f)**</u>:
$A = 0.162$ cm$^2$, $Q = 1.47 \times 10^{-5}$ C
Surface coverage $\Gamma = Q/(nFA) = [1.47 \times 10^{-5}$ C$] / [1 \times 9.648 \times 10^4$ C.mol$^{-1}$ x $0.162$ cm$^2] = 9.4 \times 10^{-10}$ moles.cm$^{-2} = 5.7 \times 10^{14}$ molecules.cm$^{-2}$

(ii) <u>**EG-electrode functionalized 4 mM α-naphthylacetate solution (Figure S3a)**</u>:
$A = 0.129$ cm$^2$, $Q = 1.26 \times 10^{-5}$ C
Surface coverage $\Gamma = 10.1 \times 10^{-10}$ moles.cm$^{-2} = 6.1 \times 10^{14}$ molecules.cm$^{-2}$

(iii) **HOPG functionalized with 2 mM α-naphthylacetate (Figure 1g)**:
$A = 6.11\ mm \times 5.99\ mm = 0.366\ cm^2$, $Q = 1.59 \times 10^{-5}\ C$.
Surface coverage $\Gamma = 4.5 \times 10^{-10}\ moles.cm^{-2} = 2.7 \times 10^{14}\ molecules.cm^{-2}$

(iv) **HOPG functionalized with 4 mM α-naphthylacetate (Figure 1h, S3b)**:
$A = 5.85\ mm \times 5.74\ mm = 0.336\ cm^2$, $Q = 2.63 \times 10^{-5}\ C$
Surface coverage $\Gamma = 8.1 \times 10^{-10}\ moles.cm^{-2} = 4.9 \times 10^{14}\ molecules.cm^{-2}$

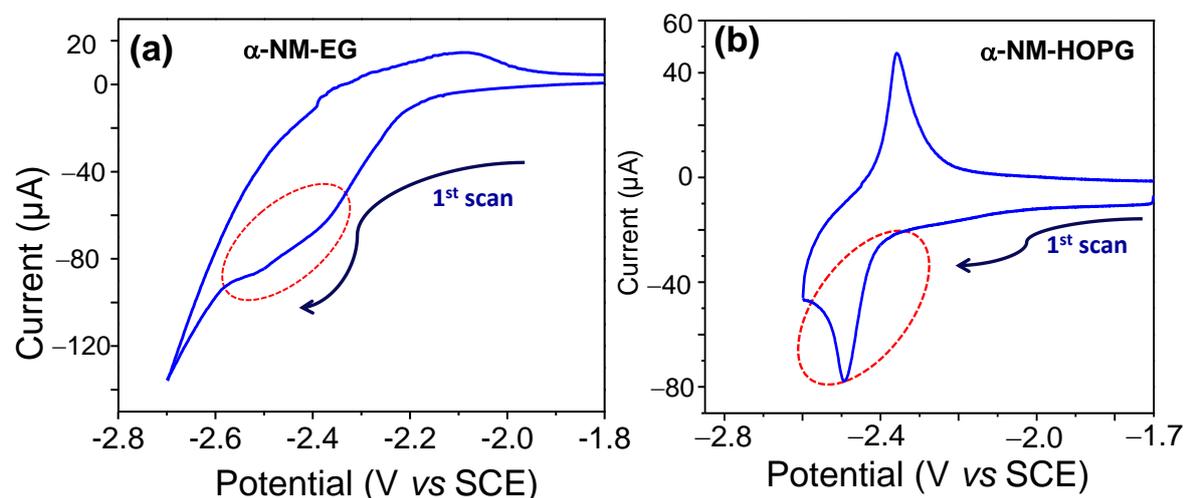

*Figure S3.* Reductive cyclic voltammograms of α−NM grafted (a) epitaxial graphene (α−NM-EG) and (b) HOPG (α−NM-EG) electrodes (derivatized as shown in **Figure S2** using 4 mM α−naphthylacetate until complete passivation of the electrode) in pure electrolyte (~0.1 M *n*-Bu$_4$NPF$_6$ in acetonitrile) solution.

**Table S1.** Electrochemical functionalization of EG and HOPG with α−naphthylacetate solutions

| Substrate | Substrate area cm$^2$ | Concentration α−Naphthyl-acetate (α−NM) | Oxidative CV (grafting of α−NM) | Reductive CV (surface coverage) | Surface coverage molecules.cm$^{-2}$ |
|---|---|---|---|---|---|
| EG-1 | 0.162 | 2 mM | Figure 1b | Figure 1f | $5.7 \times 10^{14}$ |
| EG-2 | 0.129 | 4 mM | Figure S2a | Figure S3a | $6.1 \times 10^{14}$ |
| HOPG-1 | 0.366 | 2 mM | Figure 1c | Figure 1g | $2.7 \times 10^{14}$ |
| HOPG-2 | 0.336 | 4 mM | Figure 1d, Figure S2b | Figure 1h, Figure S3b | $4.9 \times 10^{14}$ |

## 5. Electrochemical erasing of α-naphthylmethyl group from α-NM-EG:

The EG and HOPG electrodes are derivatized with α-NM groups via the formation of arylmethylene C−C bonds (ArCH$_2$−C$_{graphene}$, Ar = naphthyl), and this facilitates the formation of oxidative cleavage products as a result of benzylic resonance, which is available to stabilize the derived carbocation (Scheme 1, process 2 in the manuscript); thus the C−C bonds to the basal plane of graphene are more labile in the Kolbe product as compared to those formed in 4-nitrophenyl grafted graphene (4-NO$_2$-C$_6$H$_4$−C$_{graphene}$).[S1] The facile cleavage of the α-NM groups was observed for α-NM-EG in pure electrolyte solution by holding the potential beyond the oxidation peak, which lead to the restoration of the initial properties of the pristine EG as shown in Scheme 1, processes 4 and 5 (E = 1.85 V vs SCE for 240s).

The same electrochemical erasure can be achieved by immersing the α-NM-EG (**Figure 3a**) or α-NM-EG (**Figure 3b**) electrode in a pure electrolyte solution and running the cyclic voltammetry between 1.0 to 2.5 V vs SCE at a scan rate of 0.2 Vs$^{-1}$ for 2 cycles. After electrochemical erasure, the EG or HOPG electrodes behave like a clean EG or HOPG electrode, as could be seen by running a reductive cyclic voltammetry, which are essentially featureless (**Figure 3c and 3d**). This process of electrochemical erasing can be conveniently monitored with Raman spectroscopy (**Figure 4 and Figure S6**).

## 6. Raman spectroscopy of EG and α-NM-EG:

In the Raman spectra of graphene, the G band located at ~ 1580 cm$^{-1}$, is a first order Raman effect where the energy of the scattered incident monochromatic light is proportional to the energy of quantized lattice vibrations (E$_{2g}$ phonon) created due to the scattering process.[S2-S6] On the other hand, the 2D band (located at ~2670 cm$^{-1}$, also referred sometimes as G' peak) is a second order Raman effect caused when the lattice vibrations due to the first order process activates another phonon. In case of a single-layer graphene, the 2D peak appears as a single peak and its intensity is generally higher than the intensity of G-peak, i.e. I$_{2D}$/I$_G$ ≥ 1.

As a result of covalent chemical modification of graphene, which is usually accompanied by conversion of sp$^2$ hybridized carbons to sp$^3$, the A$_{1g}$ breathing vibration mode is activated and this results in the formation of a D-peak located at ~1345 cm$^{-1}$; D-peaks can also be seen in physically defective graphite materials, graphene nanoribbons (GNRs), edges of graphene, and disordered graphene samples. Covalent grafting of α-naphthylmethyl group to epitaxial graphene resulted in the appearance of a sharp D peak at ~1321 cm$^{-1}$ (with I$_D$/I$_G$ = 1.1) and the D′ peak at ~1610 cm$^{-1}$ (very weak) appears next to the G peak (**Figure S4b**).

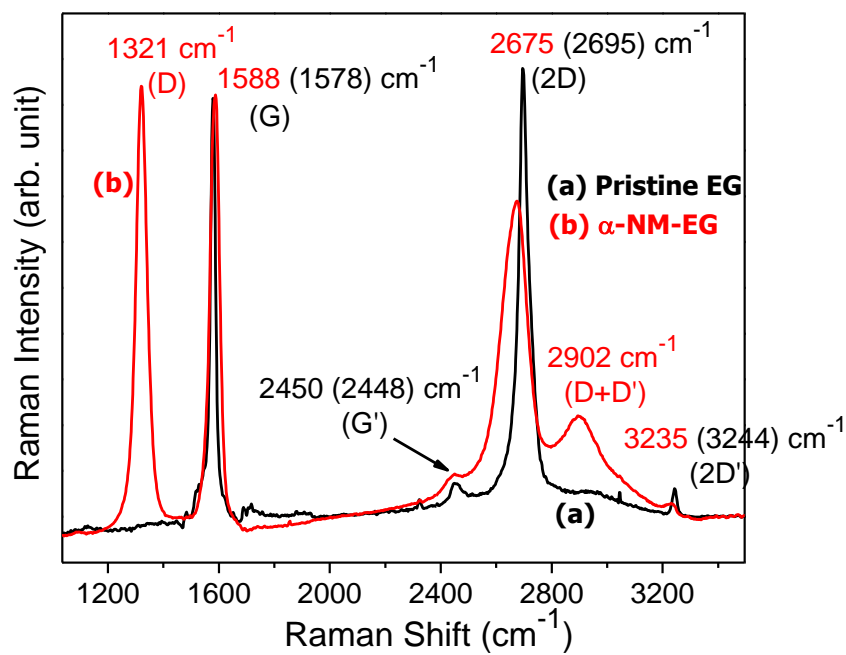

**Figure S4.** Raman spectroscopic (λ$_{ex}$ = 532 nm and spot size = 0.7 um) signatures of covalent grafting of α-NM functionality to EG-surface. (a) Pristine EG (before functionalization) and (b) after functionalization (α-NM-EG). The spectra are normalized to have similar intensity of the G peak. Raman shifts: α-NM-EG, outside bracket (pristine EG, inside bracket).

The 2D peak (~2675 cm$^{-1}$) becomes slightly broadened and its intensity decreases relative to the G peak (~1588 cm$^{-1}$). The combination mode (D+D′) at ~2840 cm$^{-1}$ in α-NM-EG, which unlike 2D and 2D′ bands requires a defect for its activation because it requires a combination of two phonons with different momentum,[S7] is absent in the pristine EG (**Figure S4a**).

## 7. Comparison of Raman spectra of pristine epitaxial graphene (EG), α-naphthylmethyl grafted EG (α-NM-EG), p-nitrophenyl grafted EG (NP-EG), and graphene oxide (GO):

**Figure S5** compares the Raman spectrum of α-NM-EG with *p*-nitrophenyl grafted epitaxial graphene (NP-EG)[S5,S6] and graphene oxide (GO).[S8] The D peak in α-NM-EG (**Figure S5b**) is observed at 1321 cm$^{-1}$, with its full-width at half maximum (FWHM) ~45 cm$^{-1}$.

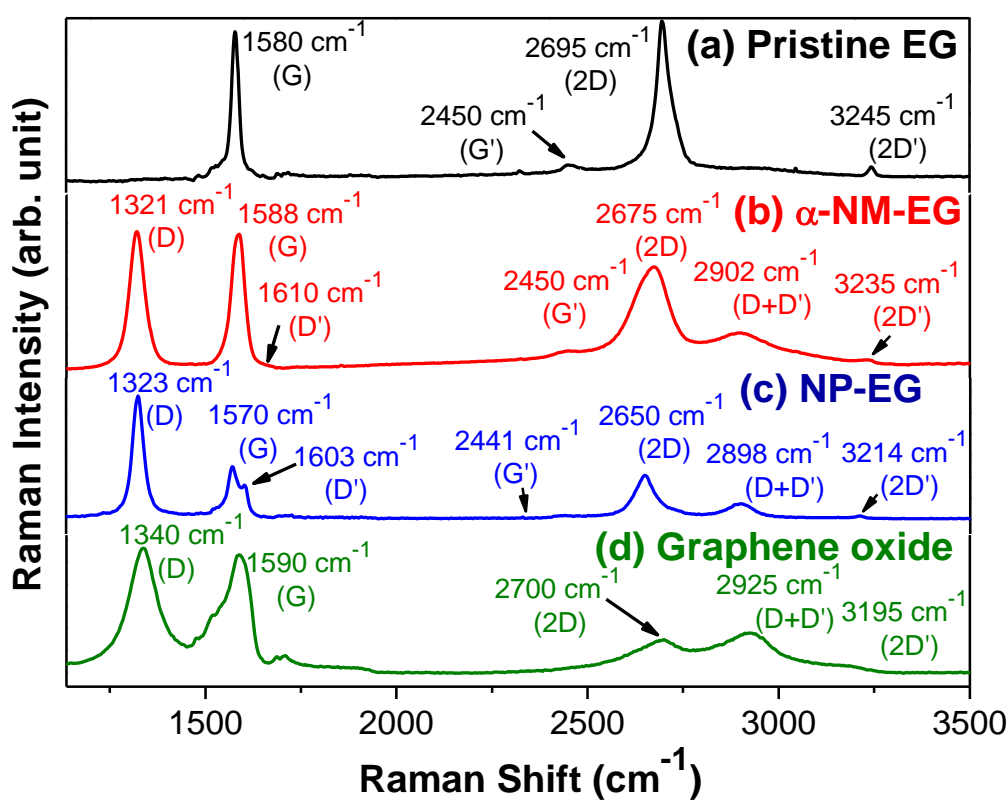

**Figure S5.** Comparison of Raman spectra ($\lambda_{ex}$ = 532 nm and spot size = 0.7 μm) of (a) pristine EG, (b) α-NM-EG, (c) NP-EG, and (d) graphene oxide (GO).

This behavior is similar to that observed in studies of 4-nitrophenyl functionalized epitaxial graphene, where the D peak appears at ~1330 cm$^{-1}$, with a FWHM of 35 cm$^{-1}$ (**Figure S5c**) and is much sharper than that observed in strongly disordered graphene oxide (GO), where the D peak appears at ~1340 cm$^{-1}$, with FWHM of 90 cm$^{-1}$ (**Figure S5d**). It should be mentioned that for graphene oxide, the G and D peaks are of comparable intensity and the spectrum shows a smaller 2D band (unlike α-NM-EG and NP-EG, where the 2D peak is very intense).

## 8. Evolution of Raman spectra under multiple electrochemical grafting and erasin:

Raman spectroscopy was found to be particularly helpful in monitoring the changes induced by multiple grafting and electro-erasing steps. This is because the electrochemical grafting of α-naphthylmethyl functionality to EG via the formation of C–C single bonds leads to the creation of $sp^3$ carbon centers at the cost of $sp^2$ centers in the graphene lattice. The electrochemical grafting of α-NM group to epitaxial graphene leads to a strong increase in the D-peak intensity, whereas the electrochemical erasing of the α-NM functionality from α-NM-EG leads to restoration of the Raman features of the pristine state as illustrated in **Figure 4c.**

Based on this Raman spectroscopic data we plotted the ratios of relative intensities of D-band to G band ($I_D/I_G$) against the repeated grafting and de-grafting processes of a selected epitaxial graphene sample (**Figure S6**).

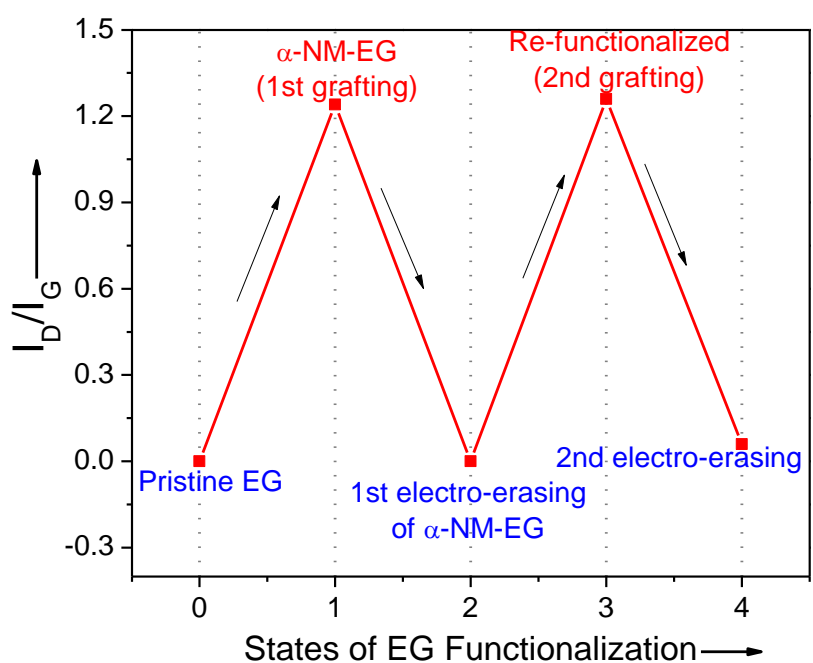

**Figure S6.** Raman demonstration of the reversibility of the grafting process; the $I_D/I_G$ ratios are plotted against the functionalization states of EG (in X-axis: 0 corresponds to pristine EG state, 1 corresponds to first grafting of α-NM, 2 corresponds to electro-erased α-NM-EG, 3 corresponds to second grafting of α-NM to electro-erased α-NM-EG, and 4 corresponds to electro-erasing of the re-functionalized α-NM-EG).

## 9. Comparision of ATR-IR spectra of epitaxial graphene, α-naphthylacetic acid and α-NM-EG:

ATR-IR data obtained for the α-NM-EG (**Figure 2d-III**) is in good agreement with the literature values for α-substituted naphthalene compounds (below, in $cm^{-1}$): 792 (in-phase C–H wagging vibrations of aryls, most intense band), 1420 ($sp^2$ C–H aromatic, bending), 1456 ($sp^3$ $CH_2$ bending), 1490 ($sp^3$ $CH_2$, asymmetric, bending), 1508 ("in-plane" C-H ring band), 1529 (C=C bend, aromatic), 1575 (C=C stretch, graphene), 1636 (C=C, stretch), 1648 (C=C, stretch, alkenyl), 2853 ($sp^3$, $CH_2$, C–H, symmetric stretch), 2924 ($sp^3$, $CH_2$, C-H, asymmetric stretch), and 3020 (sp$^2$, C–H stretch, aromatic).[S9, S10] For α-naphthylacetic acid (**Figure 2d-II**) the following characteristics IR peaks are observed: 778 cm$^{-1}$ (in-phase C–H wagging vibrations of aryls, most intense band), 1220 cm$^{-1}$ (C–O stretch, -COOH), 1410 cm$^{-1}$ (–C–H bend, alkyls), 1690 cm$^{-1}$ (C=O, stretch of COOH, usually at 1700-1725 cm$^{-1}$).[S10]

**Comparison of the low-frequency region of FTIR spectra of α-naphthylacetic acid, naphthalene and α-NM-EG:** Aryl C–H wagging bands in the IR spectra of polyaromatic hydrocarbons are usually very strong and characteristic of the aryl groups.[S9, S10] In the ATR-IR spectra of α-naphthylacetic acid, the aryl C–H wagging band appears at 778 cm$^{-1}$ (**Figure S7a**), while in naphthalene it appears at 774 cm$^{-1}$ (**Figure S7b**). The same C–H wagging vibrations of α-NM-EG appear at 792 cm$^{-1}$ (**Figure S7c**).

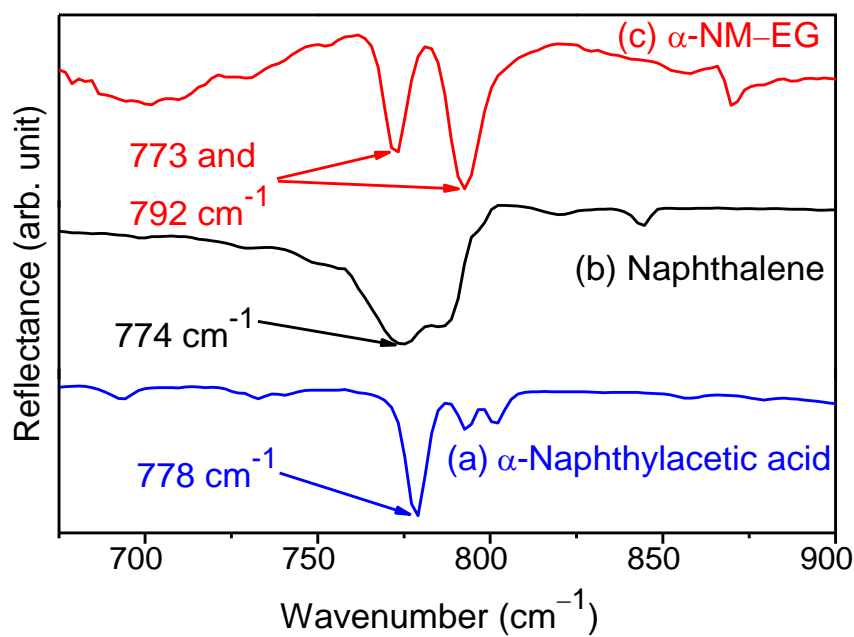

**Figure S7.** Low-frequency ATR-IR spectroscopy of (a) α-naphthylacetic acid, (b) naphthalene, and (c) α-naphthylmethyl (NM) grafted EG, showing the aryl C–H wagging bands.

## 10. Controlled electro-oxidation of EG:

In order to prepare α-NM-EG substrates with low surface coverage of the α-naphthylmethyl groups we conducted a potentiostatic electrolysis of the EG electrode for 2.5 seconds at 0.8 V vs SCE (**Figure S8a**) in presence of ~2.0 mM α-naphthylacetates (with ~0.1 M $n$-Bu$_4$NPF$_6$ in CH$_3$CN). The subsequently recorded reductive cyclic voltammogram shows the reduction peak at a potential ~2.5 V vs SCE (**Figure S8c**, after cleaning the functionalized EG electrode repeatedly with acetone, ethanol, and dichloromethane, drying with gentle flow of argon, and then transferring to a pure electrolyte solution of ~0.1 M $n$-Bu$_4$NPF$_6$ in CH$_3$CN). The integrated area (**Figure S8d**) of this reduction peak corresponds to a charge, Q = 7.62 × 10$^{-7}$ Coulombs. This corresponds to a surface coverage of ~0.49 × 10$^{-10}$ mol.cm$^{-2}$.

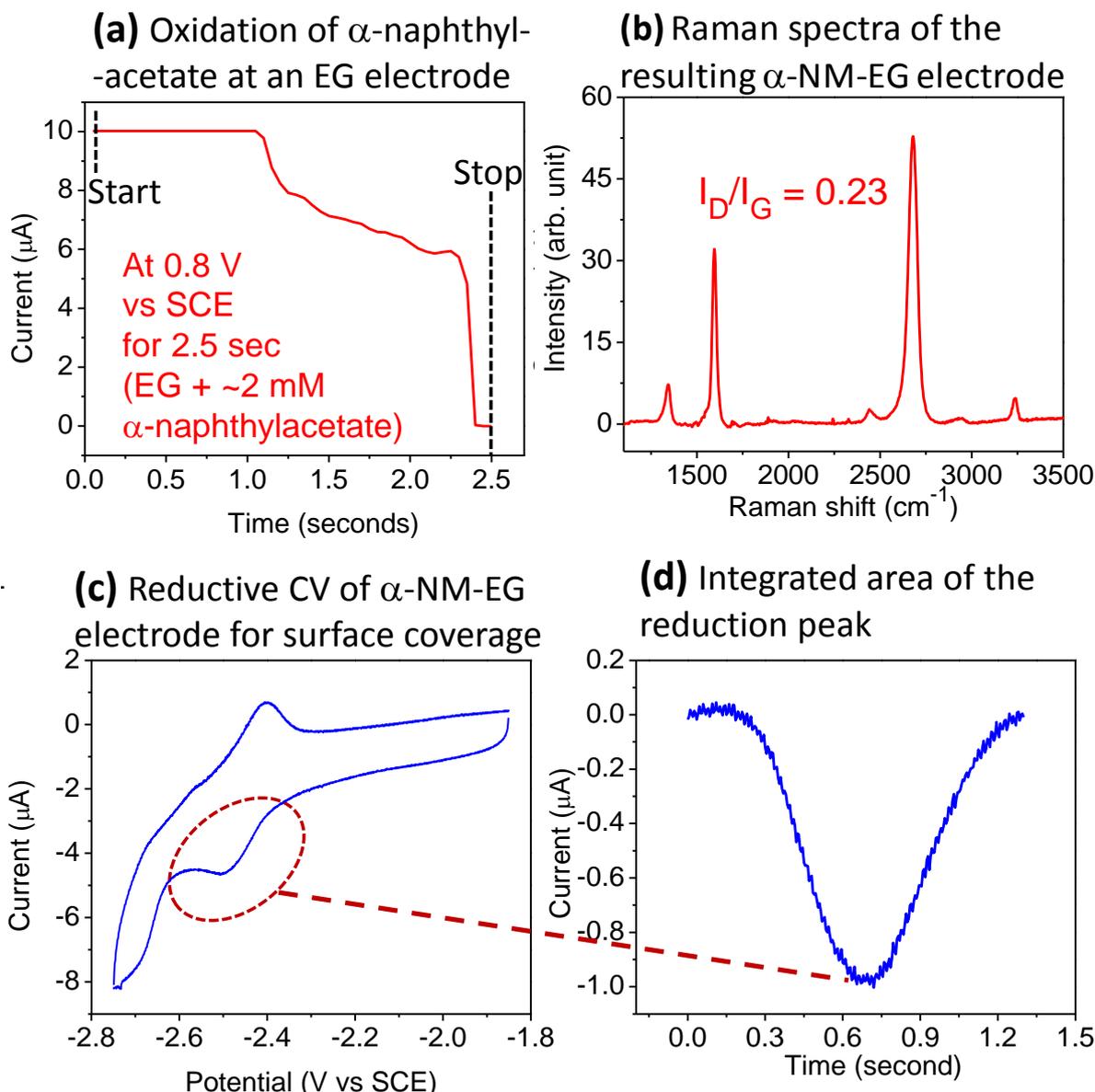

**Figure S8.** (a) Potentiostatic electrolysis of ~2 mM α-naphthylacetate at 0.8 V vs SCE on a pristine EG (working electrode) in ~0.1 M n-Bu$_4$NPF$_6$ in CH$_3$CN for 2.5 seconds at a scan rate of 0.2 V.s$^{-1}$. (b) Raman spectra (λ$_{ex}$ = 532 nm), which shows the evolution of the weak D-band at about 1345 cm$^{-1}$ (with I$_D$/I$_G$ = 0.23) in the resulting α-NM-EG electrode (**EG$_c$**), derivatized using the above method. (c) Reductive CV of α-NM-EG in a 0.1 M acetonitrile solution of n-Bu$_4$NPF$_6$, and (d) baseline corrected reduction peak of α-NM-EG at 2.5 eV *vs* SCE.

## *Supporting references*